\def\ts     {\thinspace}
\def\kms    {\ifmmode{{\rm \ts km\ts s}^{-1}}\else{\ts km\ts s$^{-1}$}\fi}
\def\msol   {\ifmmode{{\rm M}_{\odot} }\else{M$_{\odot}$}\fi}
\def\lsol   {\ifmmode{L_{\odot}}\else{$L_{\odot}$}\fi}
\def\lfir   {\ifmmode{L_{\rm FIR}}\else{$L_{\rm FIR}$}\fi}
\def\zsol   {\ifmmode{{\rm Z}_{\odot}}\else{Z$_{\odot}$}\fi}
\def\etal   {{\rm et\ts al.}}
\def\aco    {\ifmmode{{\rm CO}(J\!=\!1\! \to \!0)}\else{{\rm CO}($J$=1$\to$0)}\fi}
\def\bco    {\ifmmode{{\rm CO}(J\!=\!2\! \to \!1)}\else{{\rm CO}($J$=2$\to$1)}\fi}
\def\cco    {\ifmmode{{\rm CO}(J\!=\!3\! \to \!2)}\else{{\rm CO}($J$=3$\to$2)}\fi}
\def\dco    {\ifmmode{{\rm CO}(J\!=\!4\! \to \!3)}\else{{\rm CO}($J$=4$\to$3)}\fi}
\def\eco    {\ifmmode{{\rm CO}(J\!=\!5\! \to \!4)}\else{{\rm CO}($J$=5$\to$4)}\fi}
\def\fco    {\ifmmode{{\rm CO}(J\!=\!6\! \to \!5)}\else{{\rm CO}($J$=6$\to$5)}\fi}
\def\gco    {\ifmmode{{\rm CO}(J\!=\!7\! \to \!6)}\else{{\rm CO}($J$=7$\to$6)}\fi}
\def\hco    {\ifmmode{{\rm CO}(J\!=\!8\! \to \!7)}\else{{\rm CO}($J$=8$\to$7)}\fi}
\def\ico    {\ifmmode{{\rm CO}(J\!=\!9\! \to \!8)}\else{{\rm CO}($J$=9$\to$8)}\fi}
\def\jco    {\ifmmode{{\rm CO}(J\!=\!10\! \to \!9)}\else{{\rm CO}($J$=10$\to$9)}\fi}
\def\kco    {\ifmmode{{\rm CO}(J\!=\!11\! \to \!10)}\else{{\rm CO}($J$=11$\to$10)}\fi}
\def\ci     {\ifmmode{{\rm C}{\rm \small I}}\else{C\ts {\scriptsize I}}\fi}
\def\hi     {\ifmmode{{\rm H}{\rm \small I}}\else{H\ts {\scriptsize I}}\fi}
\def\hh     {\ifmmode{{\rm H}_2}\else{H$_2$}\fi}
\def\cone {\ifmmode{{\rm C}{\rm \small I}(^3\!P_1\!\to^3\!P_0)}
     \else{C\ts {\scriptsize I}{\small$(^3\!P_1\!\to^3\!\!\!P_0)$}}\fi}
\def\ctwo {\ifmmode{{\rm C}{\rm \small I}(^3\!P_2\!\to^3\!P_1)}
     \else{C\ts {\scriptsize I}{\small$(^3\!P_2\!\to^3\!\!\!P_1)$}}\fi}

\def\cline{\ifmmode{{\rm C}{\rm \small I}(2-1)}\else{C\ts {\scriptsize I}(2--1)}\fi}
\def\cij {\ifmmode{{\rm C}{\rm \small I}\,(^3P_i\to^3P_j)}\else{C\ts {\scriptsize I}\,{\small$(^3P_i\to^3P_j)$}}\fi}
\def\cii    {\ifmmode{{\rm C}{\rm \small II}}\else{C\ts {\scriptsize II}}\fi}
\def\tex {\ifmmode{{T}_{\rm ex}}\else{$T_{\rm ex}$}\fi}
\def\tmb {\ifmmode{{T}_{\rm mb}}\else{$T_{\rm mb}$}\fi}
\def\tkin {\ifmmode{{T}_{\rm kin}}\else{$T_{\rm kin}$}\fi}
\def\microns {\ifmmode{\mu{\rm m}}\else{$\mu$m}\fi}
\def\um{\ifmmode{\mu{\rm m}}\else{$\mu$m}\fi}
\def\nhh   {\ifmmode{n({\rm H}_2)}\else{$n$(H$_2$)}\fi}
\def\gradv {\ifmmode{{\rm dv/dr}}\else{dv/dr}\fi}
\def\rxj   {\ifmmode{{\rm RXJ0911.4+0551}}\else{RXJ0911.4+0551}\fi}
\def\dfof  {\ifmmode{{\Delta F/F}}\else{$\Delta F/F$}\fi}
\def\adfof  {\ifmmode{{\mid\!\Delta F/F\!\mid}}\else{$\mid\!\Delta F/F\!\mid$}\fi}
\def\daoa  {\ifmmode{{\Delta \alpha/\alpha}}\else{$\Delta \alpha/\alpha$}\fi}
\def\adaoa  {\ifmmode{{\mid\!\Delta \alpha/\alpha\!\mid}}\else{$\mid\!\Delta \alpha/\alpha\!\mid$}\fi}
\def\aadot  {\ifmmode{{\mid\!\dot{\alpha}/\alpha\!\mid}}\else{$\mid\!\dot{\alpha}/\alpha\!\mid$}\fi}
\def\duou  {\ifmmode{{\Delta \mu/\mu}}\else{$\Delta \mu/\mu$}\fi}
\def\aduou  {\ifmmode{{\mid\!\Delta \mu/\mu\!\mid}}\else{$\mid\!\Delta \mu/\mu\!\mid$}\fi}

\documentclass{emulateapj}

\slugcomment{accepted by ApJ 24 Apr. 2012}

\shorttitle{Variations of the Fundamental Constants at z=2.79}
\shortauthors{Wei\ss, A.\ et al.}

\begin{document}

\title{On the Variations of Fundamental Constants and AGN feedback in
  the QSO host galaxy RXJ0911.4+0551 at z=2.79}

\author{A.\ Wei\ss\altaffilmark{1}, F.\ Walter\altaffilmark{2}, D.\
  Downes\altaffilmark{3}, C. L.\ Carilli\altaffilmark{4} , C.\
  Henkel\altaffilmark{1,5}, K. M.\ Menten\altaffilmark{1}, P.\
  Cox\altaffilmark{3}}

\altaffiltext{1}{Max-Planck Institut f\"ur Radioastronomie, Auf dem H\"ugel 69, 53121 Bonn, Germany}
\altaffiltext{2}{Max-Planck Institut f\"ur Astronomie, K\"onigstuhl 17, 69117 Heidelberg, Germany}
\altaffiltext{3}{Institut de Radio Astronomie Millimetrique, 300 Rue de la Piscine, Domaine Universitaire, 
38406 Saint Martin d'H\'eres, France}
\altaffiltext{4}{National Radio Astronomy Observartory, P.O. Box O, Socorro, NM 87801, USA}
\altaffiltext{5}{Astronomy Department, King Abdulaziz University, P.O. Box 80203, Jeddah 21589, Saudi Arabia}

\begin{abstract}

  We report on sensitive observations of the \gco\ and \ctwo\
  transitions in the z=2.79 QSO host galaxy \rxj\ using the IRAM
  Plateau de Bure interferometer (PdBI).  Our extremely high signal to
  noise spectra combined with the narrow CO line width of this source
  (FWHM = 120\,\kms) allows us to estimate
  sensitive limits on the space-time variations of the fundamental
  constants using two emission lines.  Our observations show that the
  \ci\ and CO line shapes are in good agreement with each other but
  that the \ci\ line profile is of order 10\% narrower, presumably due
  to the lower opacity in the latter line.  Both lines show faint
  wings with velocities up to $\pm250$\,\kms, indicative of a
  molecular outflow. As such the data provide direct evidence for
  negative feedback in the molecular gas phase at high redshift. Our
  observations allow us to determine the observed frequencies of both
  transitions with so far unmatched accuracy at high redshift.  The
  redshift difference between the CO and \ci\ lines is sensitive to
  variations of \dfof\ with $F=\alpha^2/\mu$ where $\alpha$ is the
  fine structure constant and $\mu$ the proton-to-electron mass ratio.
  We find $\dfof=6.9\pm3.7\times 10^{-6}$ at a lookback time of
  11.3\,Gyr, which within the uncertainties, is consistent with no
  variations of the fundamental constants.

\end{abstract}


\section{Introduction} 
Possible time variation of coupling strengths and elementary particle
masses is now being discussed with regard to the accelerating,
expanding Universe.  Theoretical models that impose extra dimensions
predict that dimensionless quantities like the fine-structure constant
$\alpha=e^2/\hbar c$ or the proton-to-electron mass ratio
$\mu=m_p/m_e$, depend on the scale length of extra dimensions in
Kaluza-Klein, superstring theories or other grand unification theories
\citep[see, e.g., reviews by][]{flambaum08,chiba11}.  This scale
factor may vary with cosmic time which in turn gives rise to
variations of fundamental constants in the 4D or extra-D space-time.
Very different time dependencies, from linear to slow-rolling to
oscillating variations are considered in some theoretical models
\citep[e.g.][]{marciano84,fujii00}. Such variations are now
intensively searched for both in astrophysical observations and in
laboratory experiments.

So far, essentially all sensitive astrophysical experiments that probe
variations of the fundamental constants at high redshift have used
intervening absorbing systems towards QSOs by comparing the redshifts
of different lines. This holds for optical but also for mm and cm
(\hi, OH) absorption lines \citep[e.g.][]{carilli00,murphy03,henkel09,kanekar10}
which typically give rise to narrow ($\sim$\,10\,\kms\ wide) absorption features 
which in principle allow very precise estimates of the line centroids.
A possible systematic frequency shift in these experiments can arise
from the different velocity distributions of different species in the
absorbing gas \citep[the so called Doppler noise, see
e.g.][]{kozlov09}.  This is due to the fact that absorption
experiments only probe pencil beams through the interstellar medium
(ISM) of intervening galaxies which makes these experiments sensitive
to the physical and chemical small scale structure of the ISM.
This effect can be largely reduced by using multiplets from a single 
species \citep[e.g.][]{webb99} and large samples of absorbing systems that are now
available in the optical/UV \citep[e.g.][]{king12}.\\
For absorption studies in the radio regime the analysis can be further 
complicated by the fact that the continuum source itself may have a frequency and/or
time dependent structure (e.g. synchrotron emission from compact radio
jets) and therefore absorption may trace different parts of the
absorbing cloud. In addition such studies until today 
are limited to a few suitable objects which compromises statistical approaches. 
On the other hand radio and mm studies are particularly
sensitive probes for variations of the fundamental constants as coupling strength 
of fine-structure or inversion transitions to variations in $\alpha$ and/or $\mu$ 
is more than an order of magnitude larger than that of electronic transitions in the 
optical/UV \citep[e.g.][]{flambaum07}.

Consequently, it has been suggested to use low-- or mid--J rotational
CO lines in conjunction with fine structure lines of atomic carbon to search
for variations of the fundamental constants \citep[][]{levshakov08}. 
In contrast to the absorption studies, both lines can 
be seen in emission and therefore their line profiles probe the gas distribution
and kinematics on galactic scales rather than random motion along a
single line of sight.  As such their line centroids will not depend on
the small scale physical and chemical structure of the ISM which makes
both emission lines promising candidates to complement absorption line
studies. This line combination is sensitive to variations of $F=
\alpha^2/\mu$. Variations of this quantity are related to the observed
redshift of both lines via $\Delta\,F/F = \Delta\,z/(z+1)$ where
$\Delta\,z$ is the redshift difference between the CO and \ci\ lines.
As such their sensitivity to $\alpha$-variations is about 30 times higher
than any UV line \citep[see][]{levshakov08}. The downside of using emission
lines is that the intrinsic line widths are large compared to the
absorption lines which reduces the accuracy to determine the line
centroids at the same signal--to--noise ratio.

CO and \ci\ observations of high-z galaxies are becoming available in
increasing number \citep[see][and references therein]{walter11}. These
spectra, however, typically have low velocity resolution and low
signal-to-noise and therefore cannot be used to analyze the variations
of the fundamental constants \citep{curran11}.  Furthermore most
high-z galaxies detected in \ci\ so far have large line width
($>200\,\kms$) which implies that even with high signal to noise
observations it would be extremely challenging to study \dfof\ as this
would require redshift accuracies below 1\,\kms.

In this paper we present sensitive observations of the \gco\ and
\ctwo\ transition (CO(7--6) and \cline\ hereafter) in the high-z QSO
host galaxy \rxj\ at redshift z=2.79. \rxj\ is currently the  most
suitable candidate for high-precision CO and \ci\ redshift
measurements as it exhibits very strong line emission due to
gravitational lensing and has the narrowest CO line width (FWHM=120\,\kms)
 of any high--redshift source observed so far.  
Furthermore the increased
bandwidth of millimeter facilities implies that the redshifted
CO(7--6) and \cline\ lines can now be measured simultaneously in a
single observational setup which eliminates efficiently potential
instrumental systematics.

Measuring possible changes in the fundamental constants clearly
requires very high signal--to--noise observations of the CO and \ci\
lines.  Such high quality observations have also the potential to reveal
outflows, i.e.  negative feedback from star formation and/or AGN
winds. AGN driven molecular outflows have been recently observed in
nearby active galaxies \citep{feruglio10,alatalo11,chung11,sturm11}
and are a crucial ingredient to galaxy evolution models to explain the
number density of massive galaxies and their old stellar populations
\citep[e.g.][]{matteo05}.

\section{Observations \label{observations}}

We observed \rxj\ using the IRAM PdBI in October 2011 in the compact D
configuration with 6 antennas. The receivers were tuned to 213.1\,GHz
and data were recorded using the 3.6\,GHz WideX correlator at PdBI. At
a redshift of z=2.79 the observing frequencies for CO(7--6) and
\cline\ are 212.5 and 213.2\,GHz, respectively and comfortably fall
into the bandwidth afforded by the correlator. WideX's channel
separation is 1.95\,MHz (2.75\,\kms\ at the observed frequency).
Observations were done during good observing conditions using CRL618
as primary flux calibrator and the nearby source J0906+015 as phase
and secondary amplitude calibrator. \rxj\ was observed for $\approx
15$ hours on-source resulting in a noise level of 2.2\,mJy beam$^{-1}$
at WideX's original frequency resolution (1.95\,MHz). The synthesized
beam is 2.0$''\times\,1.4''$ with a position angle of PA=10$^\circ$.
Data reduction was performed using the {\it GILDAS} package with
standard calibration, flagging and imaging steps.

\begin{figure*}[htb]
\centering
\includegraphics[height=6.7cm,angle=0]{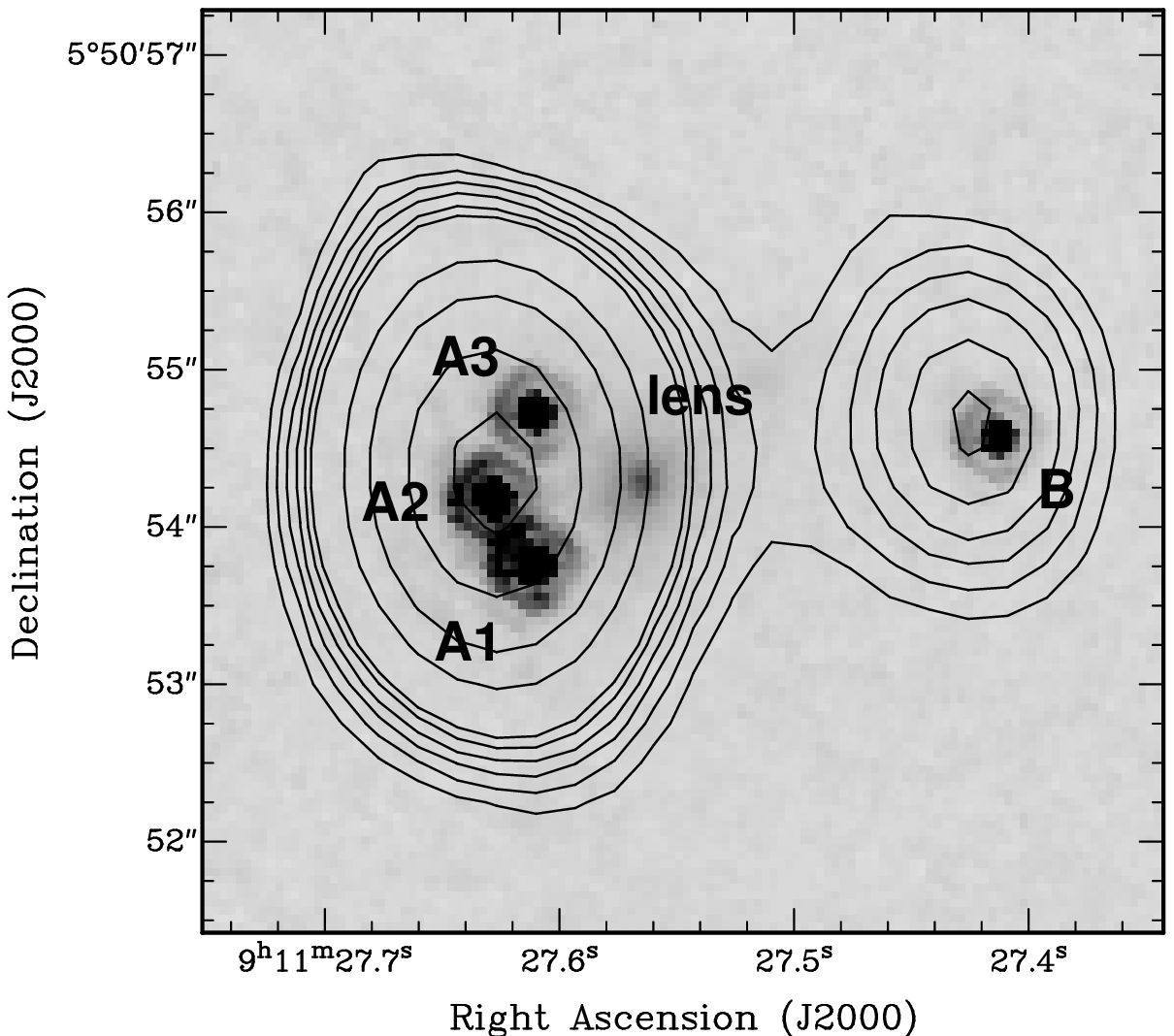}
\hspace{0.5cm}
\includegraphics[height=7.0cm,angle=0]{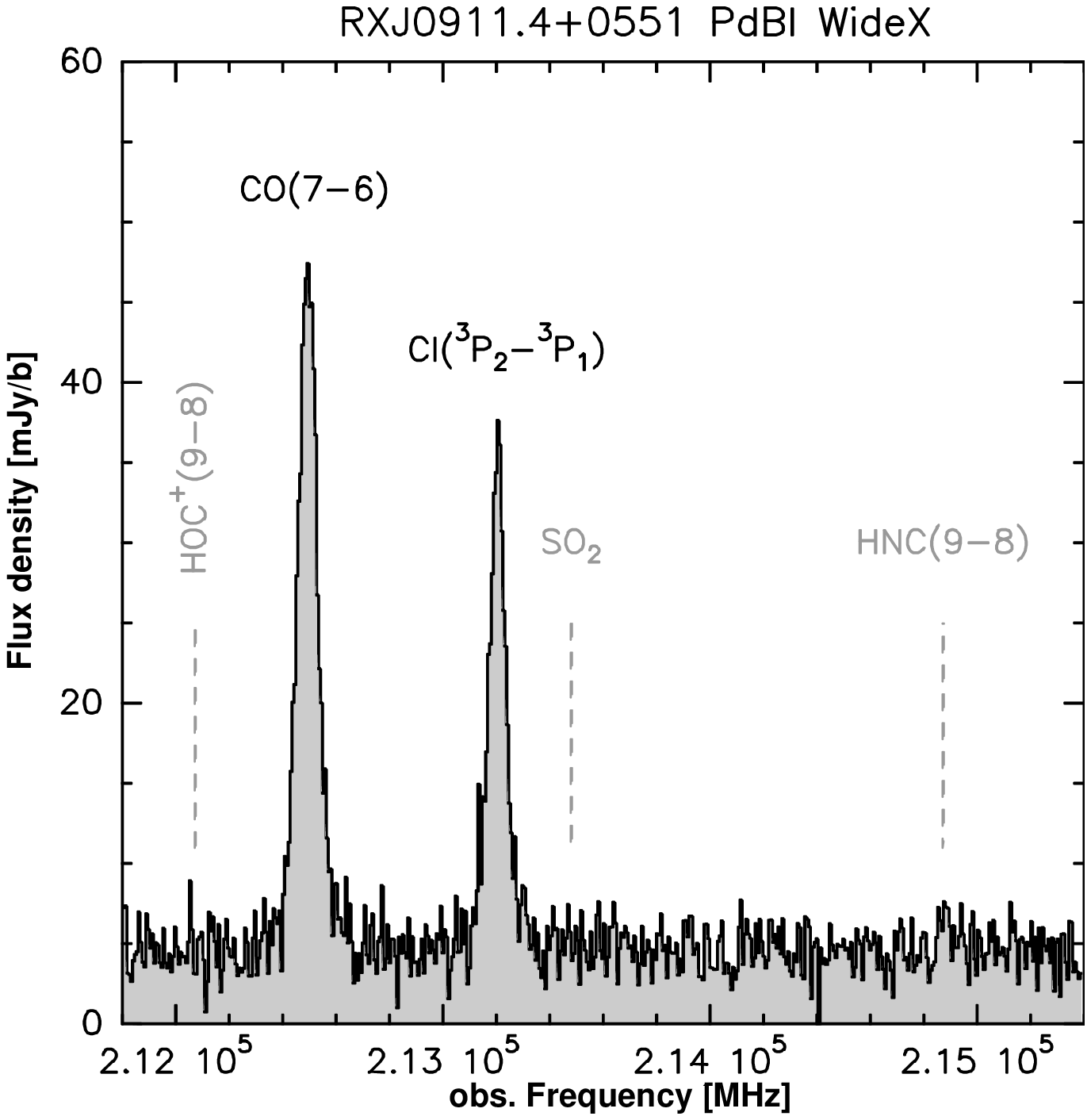}
\caption{{\it Left:} Integrated CO(7--6) intensity + continuum
  distribution (black contours) shown on an archival HST-NICMOS NIR
  image of \rxj.  Contours are shown for 0.15, 0.2, 0.25, 0.3, 0.35, 0.4, 0.7, 1.0, 1.5 and  2.0 \,Jy\,beam$^{-1}$\,\kms.   
  The eastern PdBI image (A)
  is a blend of three images \citep[A1, A2 \& A3,][]{burud98} of the
  QSO, the western peak (B) is the fourth lens image. {\it Right:}
  3.6\,GHz wide spectrum towards the peak of image A at 10\,\kms\
  resolution. Other redshifted molecular transitions which fall into the
  observing band and have been detected towards Orion \citep{comito05}
  are labeled in the figure.}
\label{result_1}
\end{figure*}

\section{Results} 
\subsection{Line and continuum distribution}

The line and continuum distribution of our observations is shown in
comparison to the near infra-red (NIR, archival HST-NICMOS image) in
Fig.\,\ref{result_1} (left).  The PdBI image shows two separate
sources. The eastern source (A) is a blend of the QSO images A1, A2 \&
A3 \citep[following the labeling by][]{burud98}, the western source is
the fourth lens image of the QSO (image B) which is spatially
separated from component A by $\approx 3''$. The NIR and mm emission regions are 
displaced with respect to each other by about $\sim0.25''$. This shift is 
obvious for image B but can also be established for A if one convolved the 
HST-NICMOS image to the PdBI beam. We interpret the offset as inaccurate
astrometry (rather than an indication for differential magnification
between the mm and NIR light), presumably in the HST-NICMOS image, as
the coordinates in the PdBI observations are phase-referenced. The
PdBI position for image B is R.A.: 09$^h$11$^m$27$^s$.430 ($\pm
0^s.01$), Dec: 05$^\circ$50$'$54.7$''$ ($\pm 0.15''$) (J2000).

The line intensity ratio between \ci\ and CO is comparable for both
components with $I_{\rm CI(2-1)}/I_{\rm CO(7-6)}=0.60\pm0.02$ and
$0.70\pm0.07$ for A and B, respectively. The center velocities (as
measured from CO) of images A and B are in good agreement and do not
show a redshift difference within the uncertainties ($\Delta
v=-4.6\pm3.1$\,\kms).  The intensity ratio between images B and A as
measured from the integrated CO(7--6) flux density is $I_{\rm
  B}/I_{\rm A}=0.21\pm0.01$, in agreement with the flux ratio measured
in the optical U-band \citep[restframe LUV
($\sim100$\,nm),][]{burud98}.  This suggests that the submm light has
a magnification similar to the optical light and in the following we
adopt the lens model by Burud \etal\ (1998) with a magnification of
$\mu_{mag}=20$.

The 3.6\,GHz wide spectrum towards image A is shown at a velocity
resolution of 10\,\kms\ in Fig.\,\ref{result_1} (right).  Several
other redshifted molecular transitions fall into the observed bandwidth. Some
lines (indicated in the figure) might have been strong enough to be
detected in \rxj\ based on the observed flux densities in the Orion-KL
hot-core \citep{comito05}. However, the only (but not significant)
additional line feature in our spectrum is the HNC(9--8) line
redshifted to 214.8\,GHz.

\subsection{Line profiles\label{sect:lineshape}}

\begin{figure*}[htb]
\centering
\includegraphics[height=5.2cm,angle=0]{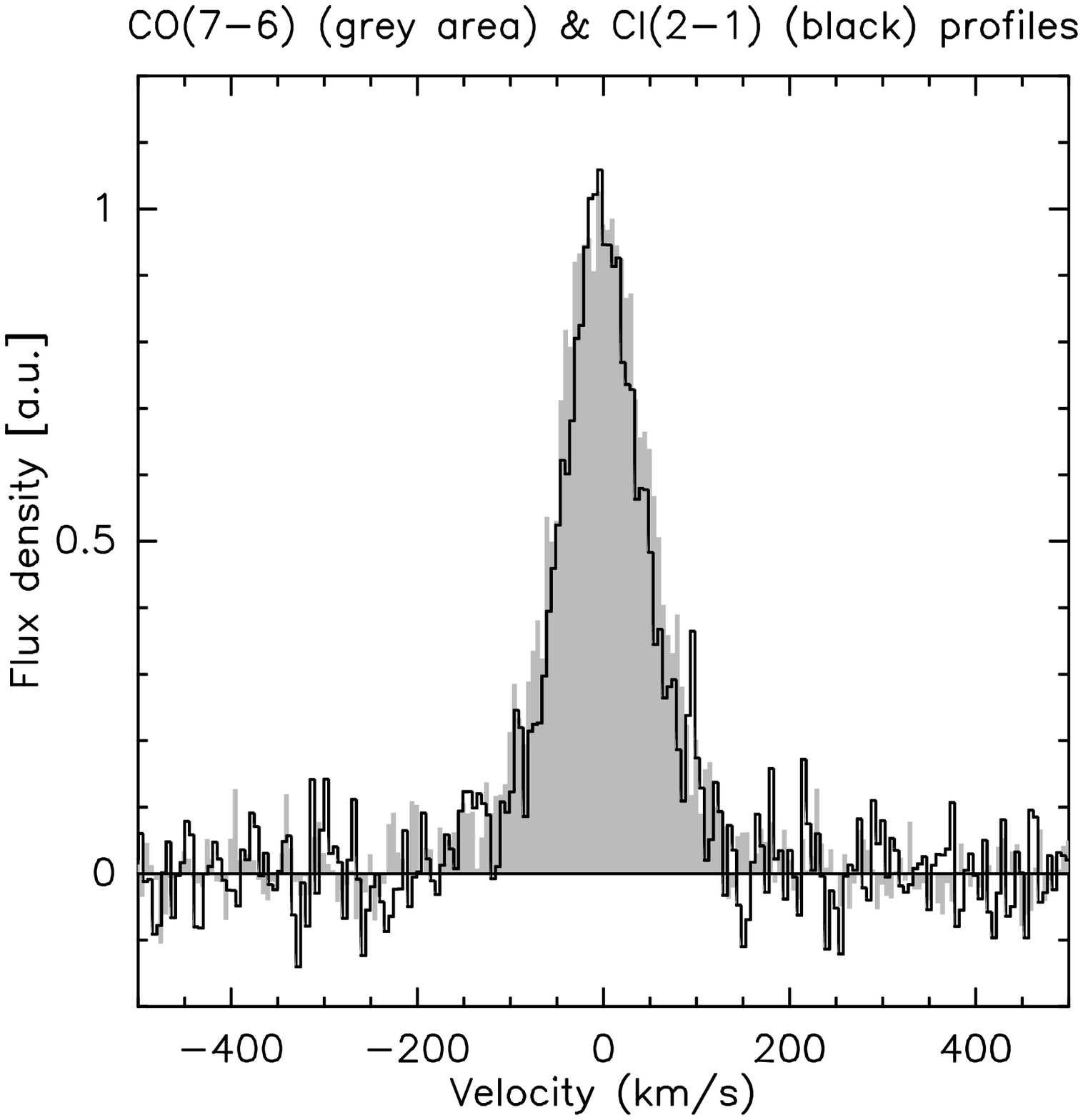}
\hspace{0.3cm}
\includegraphics[height=5.2cm,angle=0]{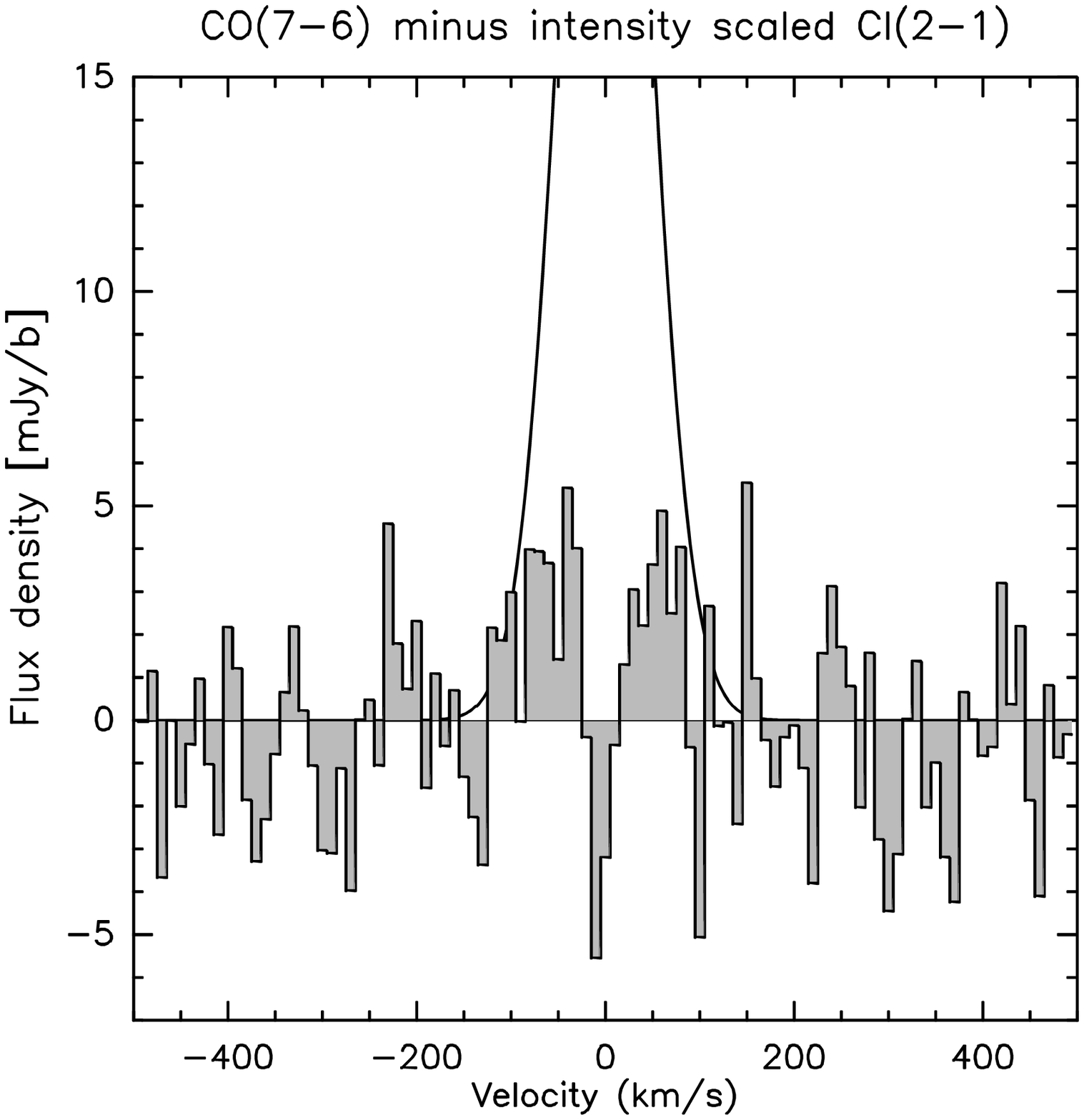}
\hspace{0.3cm}
\includegraphics[height=5.2cm,angle=0]{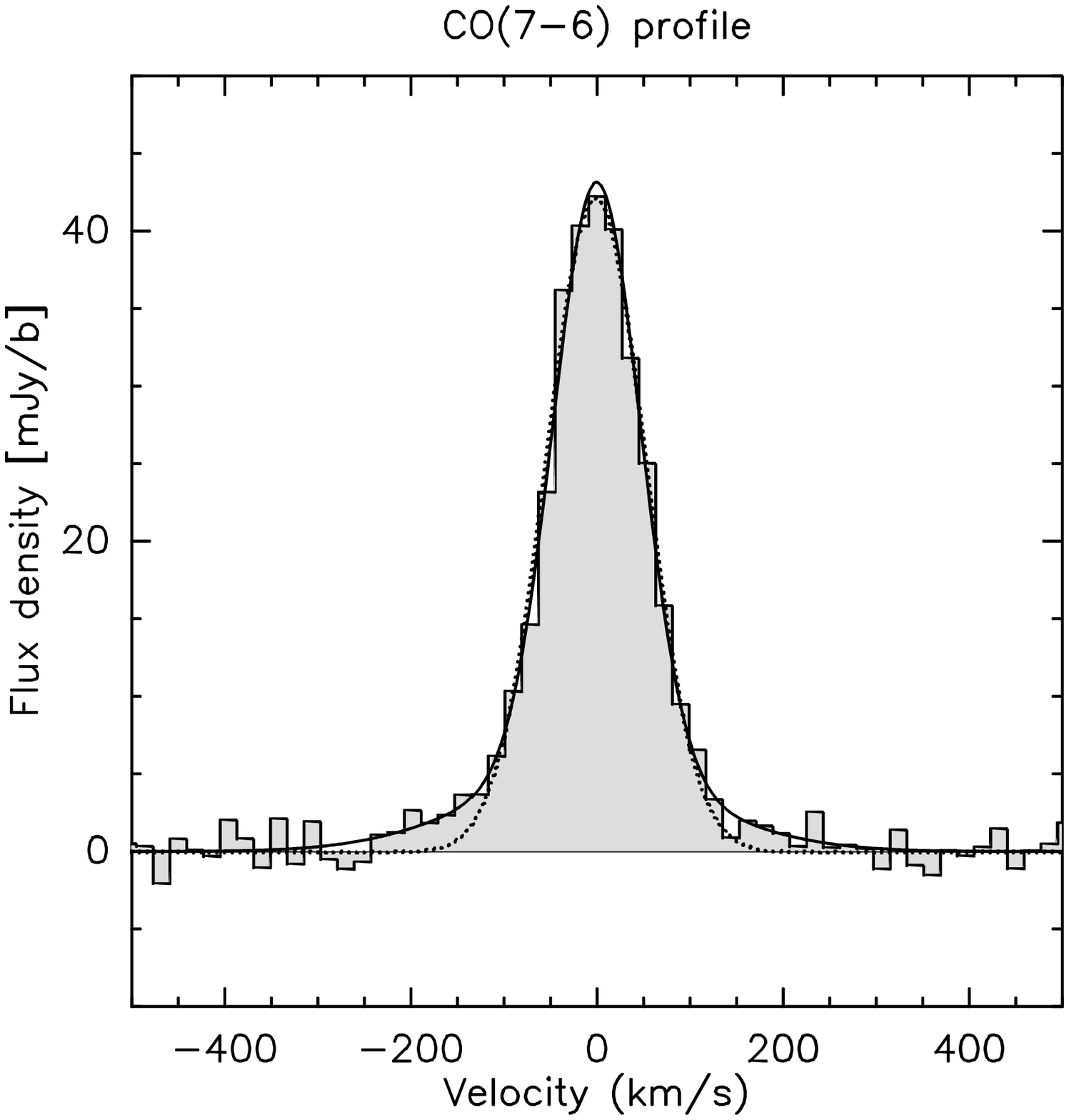}
\caption{{\it Left:} Comparison of the CO(7--6) (grey) and \cline\
  (black line) line profiles for image A at a velocity resolution of
  5\,\kms.  The velocity axis is relative to the line frequencies
  given in Table\,\ref{tab:fit}. The line intensities have
  been normalized to unity for both spectra.  {\it Middle:} Difference
  spectrum between CO and \ci\ at a velocity resolution of 10\,\kms.
  The Gaussian shows the velocity range covered by the CO line profile for
  guidance. The \ci\ intensity has been scaled to the peak of the CO
  line before subtraction. {\it Right:} CO(7--6) line profile at a
  velocity resolution of 18\,\kms.  The spectrum shows line wings that
  are detected out to $\sim\,\pm 250$\,\kms\ compared to a single
  Gaussian (dotted fit).  The black curve shows a fit with two nested
  Gaussian components which also matches the line wings.}
\label{fig:lineshape} 
\end{figure*}

A comparison of the CO(7--6) and \cline\ spectra is shown in
Fig.\,\ref{fig:lineshape} (left). Both line profiles are very similar
and almost perfectly described by a single Gaussian within the
uncertainties of the observations. A closer inspection of the two
profiles reveals that the \ci\ spectrum is slightly narrower than the
CO line. This is shown in Fig.\,\ref{fig:lineshape} (center) where we
display the difference spectrum CO--\ci\ after scaling the \ci\ peak
intensity to the peak of the CO line.  The residual shows faint but
significant CO excess in the $\mid\!50-150\!\mid \,\kms$ velocity
intervals.  From a Gaussian fit to the full line we find FWHMs of
$119 \pm 1.8\kms$ and $107\pm 2.4\kms$ for CO and \ci, respectively
(see Table\,\ref{tab:fit}).

The line profiles start to deviate from a Gaussian at 10\% level of
the peak intensity for both lines. The line wings are best seen in the
higher signal-to-noise CO(7--6) line profile at low velocity
resolution (see Fig.\,\ref{fig:lineshape}, right). The wings are also
detected in the \cline\ line albeit at lower significance. The wings
are visible both towards the red and the blue side of the spectrum and
are detected out to $\sim\,\pm 250$\,\kms. Fitting the CO(7--6) line
profile with two nested Gaussian yields an improved fit to the line
profile including the line wings. For the broad line component this 
yields a peak of $\approx 4.5$\,mJy and a line width of FWHM$\approx290$\,\kms.

\subsection{Line centroids and limits on the fundamental constants\label{sect:linecenter}}

To determine the line centroids we fitted the line shape by models with a 
single Gaussian and two nested Gaussians.
The line centroids for both methods agree well within the uncertainties
even when all 6 parameters of the nested Gaussians are kept free. The fitting 
uncertainties, however, increase by $\approx50\%$ in the latter case. We 
therefore used a single component Gaussian fit for each line and omitted the line 
wings in the fitting process. The fitting was performed on the original resolution
data as binning was found to reduce the accuracy of the results. We
have compared results from different programs (CLASS, GNUPLOT) and did
not find significant differences in the fitting results. We find
uncertainties for the line centroids of only 500 and 600\,kHz
(corresponding to a velocity uncertainty of 0.7 and 0.9\,\kms) for the
CO and \ci\ line, respectively. The results including the
corresponding redshifts are given in Table\,\ref{tab:fit}.

The redshift of the CO line ($z \equiv z_{\rm CO}$) and the redshift
difference between CO and \ci\ ($\Delta\,z \equiv z_{\rm
  CO}-z_{\ci}$), are related to the variations of the fundamental
constants via $\Delta\,z/(1+z)= \dfof$ with $F=\alpha^2/\mu$ where
$\alpha$ is the fine structure constant and $\mu$ the
proton-to-electron mass ratio \citep{levshakov08}. With the numbers in
Table\,\ref{tab:fit} we find $\Delta z=2.62\pm1.39 \times 10^{-5}$ and
thus $\dfof=6.91 \pm 3.67 \times 10^{-6}$.

\begin{deluxetable}{lclc}

\tablecaption{Derived line parameters\label{tab:fit}}
\startdata
\tableline
Line & $\nu_{\rm obs}$ & $z_{\rm line} \tablenotemark{a}$ & $FWHM$ \\
  &         [GHz]  &                & [\kms]\\

\tableline
CO(7--6)& 212.4935(5)     &2.796125(9)   & $119\pm1.8$ \\
\cline\ & 213.2036(6)     &2.796099(11)  & $107\pm2.4$ \\
\tableline
\enddata
\tablenotetext{a}{Calculated using rest frequencies of 806.6518060(5) and 809.341970(5) GHz 
CO(7--6) for  and \cline\, respectively \citep{mueller01}. }

\end{deluxetable}

\section{Discussion}

\subsection{Line wings}
The line wings detected in our spectra towards \rxj\ are remarkably
similar to those recently detected in deep CO observations towards two
local galaxies hosting an AGN: Mrk\,231 \citep{feruglio10} and
NGC\,1266 \citep{alatalo11}.  In both cases the CO wings have been
interpreted as molecular outflows. While the projected outflow
velocities in NGC\,1266 are comparable to those detected in our
spectra, the outflow signature in Mrk\,231 is much more pronounced
with wings detected out to $\pm\,750$\,\kms. Although it is tempting
to interpret the line wings in \rxj\ as the signatures of a molecular
outflow the situation is complicated by the fact that the source is
amplified by a galaxy (visible in the HST-NICMOS image shown in
Fig.\,\ref{result_1}) and only to a lesser degree by a nearby galaxy
cluster \citep[see][]{burud98,kneib00}. This implies that the caustic
curves are very compact, unlike the smoothly varying caustics from
cluster lensing.  Thus an alternative interpretation is that the lens
only strongly amplifies a small part of the molecular gas distribution
in the disk of \rxj\ (with correspondingly narrow CO line width) and
that the faint, broader CO profile visible as wings is the true line
width of the system with low or no lens magnification.

We can, however, use the observed line luminosity of the optically
thick CO(7--6) line together with the excitation temperature of
$T_{\rm ex} \approx 35$\,K estimated from the \ci\ line ratio (using
the intensity of the lower carbon line from Walter \etal\ 2011) to
obtain a lower limit for the size of the amplified emission region.
This calculation uses 
$\Omega_s=L'_{CO(7-6)}\,(T_{b\,CO(7-6)}\,d{\rm v}\,D^2_A)^{-1}\,\mu_{mag}^{-1}$ 
\citep{solomon97} and assumes $T_{b\,CO(7-6)} \approx T_{ex\,C\small I}-T_{cmb}$.
Taking the magnification of $\mu_{mag}=20$ into account this yields a source
diameter of $\sim\!1$\,kpc which is a lower limit to the true source
size because 1) it assumes an area filling factor of the emission of
unity, 2) a face--on geometry, and 3) the true excitation temperature
of the CO(7--6) line is likely lower than our estimate from \ci\ due
to the higher critical density of the CO(7--6) transition compared to
both \ci\ lines. Thus the true underlying source size is likely
similar to what has been observed in resolved, unlensed high redshift
QSO host galaxies \citep[such as J1148+5251 with
D=2.5\,kpc,][]{walter04}. This comparison suggests that the
gravitational lens amplifies the entire molecular gas distribution of
\rxj\ and not just a small fraction of the disk.  This favors the
interpretation that the narrow observed line width is due to a close
to face--on orientation of the molecular disk and that the faint line
wings indeed trace a molecular outflow.

We estimate the molecular gas mass in the outflow using the optically
thin \ctwo\ line \citep[see e.g. Eq.\,2 in][]{weiss03}.
We use the \ci\ excitation temperature given above
and a carbon abundance, $[\ci]/[\hh]$, of $8\times 10^{-5}$
\citep{walter11} which yields a molecular gas mass (including a
correction of 1.36 for Helium and correcting for the magnification) of
$M_{\rm outflow} = 6 \times 10^8\,\msol$. This mass only includes the
line wings (difference between the two component and the single
component fit shown in Fig.\,\ref{fig:lineshape} right, albeit for
\cline) and corresponds to $\approx 9\%$ of the \ctwo\ line
luminosity. The total mass in the broad line component (which has been
used by Feruglio \etal\ (2010) to estimate the outflow mass in
Mrk\,231) is $1.7 \times 10^9\,\msol$.  We here, however, use the more
conservative approach to estimate the outflow properties only based on
the mass in the line wings \citep{alatalo11}. Using the
HWHM\,=\,145\,\kms\ from the Gaussian fit to the broad line component
as a lower limit for the average outflow speed (due to the unknown
outflow orientation) the kinetic energy in the outflow is ${E}_{\rm
  kin} > 1.2 \times 10^{56}$\,erg.  Thus the outflow in \rxj\ is $\sim
3 \times$ more massive but likely less energetic than the outflow in
Mrk\,231 while other key parameters ($L_{\rm FIR}$, $M_{\hh}$, $M_{\rm
  BH}$) are comparable for both sources after correcting for the
gravitational magnification.

For \rxj\ the lack of information on the size and morphology of the
outflow prevents us from estimating the dynamical time scale and the
outflow rate.  Mapping the blue and the red line wings independently
only yields an upper limit for their angular separation in the image
plane of $<0.3''$ ($<2.4$\,kpc), not sufficient to obtain a meaningful
limit on the outflow rate. If we assume that the outflow size is
comparable to the outflows observed in Mrk\,231 and NGC\,1266
(r=0.5\,kpc) the outflow rate for \rxj\ is
$\dot{M}>180$\,\msol\,yr$^{-1}$, larger than the star formation rate
\citep[SFR=140\,\msol\,yr$^{-1}$,][]{wu09} which would hint at AGN
feedback as being the main driver for the molecular outflow \citep[see
the discussion in][]{alatalo11}.  The corresponding gas depletion time
scale would be only 40\,Myr. Even if we use the observed upper limit
of $<2.4$\,kpc as the intrinsic radius for the outflow the depletion
time scale remains short ($<200$\,Myr) which implies that a
significant fraction ($>50\%$) of molecular gas will be removed from
the disk within the typical lifetime of 100\,Myr for massive
starbursts at high redshift \citep{tacconi08}.

\subsection{Limits on the fundamental constants}

A potential concern using mid-J CO and \ci\ emission lines as a probe
of variations of the fundamental constants is whether both lines trace the
same gas on galactic scales. Observations of CO and \ci\ in the Milky
Way have shown that both lines indeed probe the same molecular
material on large and small scales \citep{ikeda02,zhang07} despite the
expectations from photon dominated region (PDR) models which predict
that \ci\ should only exist in small layers surrounding the CO cloud
cores.  As discussed in the previous section, our experiment probably
probes a region of $\sim 1$\,kpc in diameter. 
We would therefore expect that any potential spatial differences in the CO and
\ci\ emission  velocities within molecular clouds would average out to zero over these extended
regions.

Given that both lines trace the same volume on large scales one would
naively expect that their line profiles are identical. Our
observations show that this is not the case and that the \ci\ line
profile is $\sim10$\% narrower than the CO profile
(Fig.\,\ref{fig:lineshape} left \& center; Table\,\ref{tab:fit}). This effect, however, is
expected as the opacity of CO is typically much larger than the \ci\
opacity (for the peak of the lines in \rxj\ we estimate opacities of
17 and 1.0 for CO(7--6) and \cline\, respectively using large velocity
gradient models).  Since the gas column density and therefore the
opacity decreases towards the wings of the lines and because the
intensity in each velocity bin scales as $(1-e^{-\tau})$, the
intensity in the wings of the \ci\ line decreases more rapidly than in
the case of CO which in turn gives rise to a narrower line profile in
\ci.  This effect, however, is not expected to be a critical
limitation for the precise determination of the line centroid (i.e.
the systemic velocity or redshift of the galaxy): the line profiles of
galaxy-averaged molecular emission lines are not only determined by
opacity and excitation effects but also by the effective source solid
angle as a function of velocity. The latter typically does not
saturate (i.e.  clouds do not overlap in velocity space and spatially
at the same time) which prevents galaxy-averaged CO lines from
saturating at their peak.  As systematic instrumental effects (e.g.
LSR corrections) cancel out in our simultaneous observations of both
lines and local oscillator variations (defining the frequency accuracy
of the spectrometer) can safely be neglected, we conclude that
systematic uncertainties are small compared with the statistical
uncertainties, i.e. our ability to determine variations of the
fundamental constants is solely limited by the signal to noise ratio and 
line width of our spectrum.

So far estimates on the variations of the fundamental constants using
high redshift emission lines have only been published in two studies
using \cii\ and CO \citep[in BR1202-0725 at z=4.7 and J1148+5251 at
z=6.4,][]{levshakov08} and \ci\ and CO \citep[in a sample of sources
at z=2.3--4.1,][]{curran11} where the measurements were taken entirely
from the literature.  These attempts yielded 3$\sigma$ limits on
\adfof\ of $<4.5 \times 10^{-4}$ and $<2.6 \times 10^{-4}$,
respectively. Our experiment improves on these limits by more than a
factor of 20 (at comparable lookback time) with \adfof\,$<1.1 \times
10^{-5}\,(3\,\sigma)$. Most absorption line studies have focused on
variations in \daoa\ or \duou. Variations of \dfof\ are related to
both these quantities via \dfof=2\,\daoa$-$\duou\ \citep{levshakov08}.
Centimeter and millimeter absorption line studies have recently used
inversion lines of NH$_3$ in combination with rotational lines of
dense gas tracers (such as HC$_3$N or CS) to obtain limits on \duou.
This combination provides a particularly sensitive probe to \duou\ due
to the strong coupling between their redshift difference and
variations in $\mu$ \citep[e.g.][]{levshakov10}. Studies using several
inversion and rotational transitions toward PKS\,1810-211 at $z=0.89$
give a conservative limit of $\aduou < 10^{-6}$
\citep[][]{murphy08,henkel09}. The formally most accurate application
of this technique to date yields a 3\,$\sigma$ limit of
$\aduou<3.6\times 10^{-7}$ at $z=0.68$ \citep{kanekar11} albeit on a
small number of spectral lines.

Limits on variations of $\alpha$ at high redshift do not reach such
accuracy to date but some studies have reported non-zero values for
\daoa\ based on QSO absorption spectra \citep{murphy03,webb11,king12}.  We
here use our results to obtain an independent limit on \daoa.  If we
assume \duou\,=\,0 between z=0 and z=2.79 (motivated by the NH$_3$
results mentioned above) our observations yield $\daoa=3.5\pm
1.8\times 10^{-6}$, consistent with zero variations of $\alpha$ at a
lookback time of 11.3\,Gyr within the uncertainties ($\aadot <
4.8\times 10^{-16}\,{\rm yr}^{-1}\ (3\,\sigma)$). As such our
experiment yields comparable accuracy on a single object as work based
on HIRES/Keck and UVES/VLT absorption line spectroscopy on QSO samples
which give typical accuracies of 1-2\,ppm \citep[1$\sigma$,][]{levshakov07,murphy08,webb11}.
Our source, \rxj, lies nearly orthogonal to the proposed 
angular dipole distribution of Webb \etal\ (2011) and King \etal\ (2012)
and thus does not provide a stringent test of that claim.

\section{Concluding remarks}

Our observations demonstrate that sensitive simultaneous millimeter
wave observations of CO(7--6) and \cline\ emission lines provide a
powerful diagnostic to investigate possible variations of the
fundamental constants at high redshift.  Our observations do not show
evidence that systematic effects are limiting the interpretation of
the data and our results are limited by the signal-to-noise
ratio of our spectra. This implies that deeper observations with even
higher spectral resolution, which are now in reach with ALMA, will
greatly improve our current limits. A potential concern of deep and
more precise observations could arise from the different optical depth
in the CO and \ci\ lines which leads to slightly different line shapes
for both transitions. Such potential limitations can be eliminated by 
observations of reasonably sized samples of high-z CO and \ci\ emitters 
with suitable line widths. Given ALMA's sensitivity even CO isotopologues such 
as $^{13}$CO will eventually become accessible with high signal--to--noise ratio 
which will further help to remove this potential limitation. In any case 
the uncertainties and potential systematics of emission line studies will be substantially 
different from those of absorption line experiments. As such the combination of 
both approaches on samples of high-z galaxies will greatly improve the 
sensitivity and reliability of measurements probing the variations 
of the fundamental constants.

Our data also demonstrate the large discovery space of deep
spectroscopic millimeter observations with large bandwidth. We detect
for the first time a massive molecular outflow via CO and \ci\ line
wings in a high redshift galaxy. Our analysis suggests that a significant
fraction ($>50\%$) of molecular gas in \rxj\ will be removed from the
disk within the typical lifetime of 100\,Myr for massive starbursts.
As such \rxj\ is one of the few examples for negative feedback
observed in the molecular gas phase at high redshift, although higher
resolution observations (e.g. with ALMA) will be required to decide whether the outflow
is driven by an AGN wind or by star formation.

\acknowledgments
We thank the referee, John Webb, for useful comments that helped to 
improve the manuscript. This work is based on observations carried out 
with the IRAM Plateau de Bure Interferometer. IRAM is supported 
by INSU/CNRS (France), MPG (Germany) and IGN (Spain).

\end{document}